# Classification of Charge Density Waves Based on Their Nature

Xuetao Zhu[1], Yanwei Cao[1], Jiandi Zhang[2], E. W. Plummer[2], and Jiandong Guo[1]

[1] *Institute of Physics, Chinese Academy of Sciences and Beijing National Laboratory for Condensed-Matter Physics, Beijing 100190, China*

[2] *Department of Physics and Astronomy, Louisiana State University, Baton Rouge, LA 70808, USA*

**Abstract**

**The concept of a Charge Density Wave (CDW) permeates much of condensed matter physics and chemistry. Conceptually, CDWs have their origin rooted in the instability of a one-dimensional system described by Peierls. The extension of this concept to reduced dimensional systems has led to the concept of Fermi surface nesting (FSN), which dictates the wave vector ($\vec{q}_{\text{CDW}}$) of the CDW and the corresponding lattice distortion. The idea is that segments of the Fermi contours are connected by $\vec{q}_{\text{CDW}}$, resulting in the effective screening of phonons inducing Kohn Anomalies in their dispersion at $\vec{q}_{\text{CDW}}$, driving a lattice restructuring at low temperatures. There is growing theoretical and experimental evidence that this picture fails in many real systems and in fact it is the momentum dependence of the electron-phonon coupling (EPC) matrix element that determines the characteristic of CDW phase ($\vec{q}_{\text{CDW}}$). Here, based on the published results for the prototypical CDW system 2H-NbSe$_2$, we show how well the $\vec{q}$-dependent EPC matrix element, but not the FSN, can describe the origin of CDW. We further demonstrate a procedure of combing electronic band and phonon (dispersion and linewidth) measurements to extract the EPC matrix element, allowing the electronic states involved in the EPC to be identified. Thus we show that a large EPC does not necessarily induce the CDW phase, with Bi$_2$Sr$_2$CaCu$_2$O$_{8+\delta}$ (Bi2212) as the example, and the charge ordered phenomena observed in various cuprates are not driven by FSN or EPC. To experimentally resolve the microscopic picture of EPC will lead to a fundamental change in the way we think about, write about, and classify Charge Density Waves.**

The phrase Charge Density Wave (CDW) was first used by Fröhlich (1, 2) but originates from Peierls description of a fundamental instability in a one-dimensional (1D) chain of atoms equally spaced by a lattice constant *a* (2). Figure 1A shows the free electron band of such a 1D chain with one electron per atomic site. The Fermi points are at $k_F = \pm \pi/2a$ and are connected by the nesting vector $q = 2k_F$. In 1930 Peierls asserted that this system is unstable, showing an electronic disturbance with the wave vector $2k_F$, changing the periodicity of the chain, and opening up a gap at the zone boundary ($k = \pi/2a$) of the new unit cell containing two atoms (1, 2). The conjecture was that the gain in electronic energy would always overwhelm the cost of restructuring the atoms (1). Consequently in the Peierls model there would be a transition from the metallic high temperature state to the insulating-dimerized ground state at a critical temperature $T_{CDW}$. Kohn (3) pointed out that there is an image of the Fermi surface in the vibrational spectrum because the zero energy electronic excitations at $2k_F$ will effectively screen any lattice motion with this wave vector. Figure 1B shows the phonon dispersion for this 1D chain at different temperatures (1). Below $T_{CDW}$ the phonon energy at $q = 2k_F$ becomes imaginary, meaning there is a new lattice structure. Above $T_{CDW}$ there is a sharp dip (Kohn anomaly) in the phonon dispersion but no static restructuring.

The question is how to extend the Peierls picture to real systems in higher dimensions (a real one dimensional system is only quasi 1D). A standard practice is to calculate the susceptibility $\chi(\vec{q},\omega)$ for a given electronic configuration (1) and to use the zero energy value of the Lindhard response function $\chi_0(\vec{q}) \equiv \chi_0(\vec{q},\omega=0)$ (see the supplemental material) to determine if the electron response can drive a Peierls phase transition — there should be a peak in the imaginary part of the response function $\text{Im}\left[\chi_0(\vec{q})\right]$ at the FSN vector $\vec{q}_{CDW}$ as well as in the real part $\text{Re}\left[\chi_0(\vec{q})\right]$, because the real part defines the stability of the system (4). Figure 1D shows the plot of $\text{Re}\left[\chi_0(\vec{q})\right]$ for a free electron system in 1D, 2D and 3D (1), respectively, where Fig. 1C displays the free-electron Fermi contour in 2D. There is a logarithmic divergence in $\text{Re}\left[\chi_0(\vec{q})\right]$ for the 1D system but no peaks in the 2D and 3D electron gases. If CDWs are created by FSN, the Fermi contour must have been distorted to create nesting as represented by $\text{Re}\left[\chi_0(\vec{q})\right]$. In a seminal paper Johannes and Mazin demonstrated that the logarithmic divergence of $\text{Re}\left[\chi_0(\vec{q})\right]$ is not robust against small deviations from perfect nesting (4). The dotted line in Fig. 1D illustrates what happens for only a 2% deviation in $k_F$ from perfect nesting and the inset shows the effect of including a small "Drude relaxation" in the calculation for the susceptibility.

If FSN is the origin of CDW as in Peierls picture, the real and imaginary parts of the susceptibility calculated using the measured band structure should both show strong peaks at $\vec{q}_{CDW}$, there should be a Kohn anomaly in all of the phonons (expect ones not allowed by symmetry) at $\vec{q}_{CDW}$ (4), and at least one of the modes must go imaginary at low temperature, inducing a lattice distortion and an electronic gap. However, Johannes and Mazin argued, "only a tiny fraction if any, of the observed charge ordering phase transition are true analogs of the Peierls instability--." (4). In fact, it has been postulated that CDW phases are dictated by the $\vec{q}$-dependent electron-phonon coupling (EPC) (4-7). Mazin *et al.* used the classical CDW materials NbSe$_2$, TaSe$_2$, and CeTe$_3$ to justify their conclusion and showed that with a proper inclusion of EPC the observed properties could be explained (4, 6, 7). It is then obvious that experimentally we should be exploring the phonon spectra in detail to determine the $\vec{q}$ dependence of EPC (8). One may argue that both FSN and EPC are forms of electron-lattice interaction, involving excitations of electrons from filled to empty states. But they should be distinguished since FSN involves elastic scattering, while EPC involves inelastic scattering from the lattice.

Although it is possible that when the dimensionality is truly reduced the CDW state will be tied intimately to FSN, it is appropriate at this stage to quote Peierls (1972) (9): "This (Peierls) instability came to me as a complete surprise when I was tidying material for my book (1955), and it took me a considerable time to convince myself that the argument was sound. It seemed of only academic significance, however, since there are no strictly one-dimensional systems in nature. I therefore did not think it worth publishing the argument, beyond a brief remark in the book, which did not even mention the logarithmic behavior." A recent theoretical work indicated that Peierls' intuition might have been correct (4). Density functional theory calculations were performed for a 1D chain of Na atoms, which should be the best representation of a perfect Peierls system. With the atoms clamped there was no sign of an electronic-only density wave, and relaxation of the ion position in the 1D chain failed to produce any distortion, *i.e.* no Peierls distortion. However if the ions were allowed to move in 2D the lowest energy state was a zigzag chain, but the electronic gap anticipated in the Peierls CDW phase was missing (4).

Experimentally a whole host of techniques must be utilized to fully characterize the properties associated with a CDW. Angle resolved photoemission spectroscopy (ARPES) can determine the band structure and the Fermi contour (10-12) that are needed to calculate the electronic susceptibility. Electron (13, 14), X-ray (15), and neutron diffraction (16, 17) measurements can determine the structure, $\vec{q}_{CDW}$ and $T_{CDW}$. Scanning tunneling microscopy (STM) and spectroscopy (18, 19) can determine the wave length of the charge modulation as well as the electronic transition temperature. Inelastic neutron (16, 17) and X-ray scattering (20) can probe the temperature dependence of the Kohn anomaly in the phonon dispersion. Transport measurements (21) can reveal the transition temperature and the metal insulator transition predicted by the Peierls model. Finally, since EPC is important, high resolution ARPES data is needed to estimate the strength of the EPC in a CDW system (22). Here we will stress the importance of the phonon dispersion and linewidth measurements are to understanding the origin of CDWs.

To illustrate the importance of EPC in the formation of a CDW state we will briefly describe the classic and the most studied CDW system, NbSe$_2$ (23), a layered quasi-2D material. This material has a hexagonal layered structure with lattice constant of 3.443Å. A CDW phase transition occurs with $q_{CDW} \cong \frac{2}{3}|\Gamma M| = 0.695 Å^{-1}$ ($\Gamma$ and M are the points at the center and boundary in the Brillouin zone (BZ), respectively, in the high temperature phase) at $T_{CDW}$ = 33.5 K (15, 17) (shown in Fig. 2A). All of the data and the modeling for this system dramatically illustrate that FSN is irrelevant and that the origin of the CDW state is tied to EPC (6, 7, 18-20). This realization should not have been a surprise since in 2004 Valla *et al.* reported that for NbSe$_2$ EPC was the dominant contribution to the renormalization of quasiparticle self-energy (22). Figure 2A shows the Fermi contour obtained from a tight binding fit to the ARPES data (10, 12) and $\vec{q}_{CDW}$ obtained from diffraction measurements (13, 14, 16, 17). Obviously there is no way to have FSN with $\vec{q}_{CDW}$! In fact the absence of FSN can be seen in the real and imaginary parts of the susceptibility $\chi_0(\vec{q})$ plotted from the experimental data in the Γ-M direction in Fig. 2B & 2C.

There is no sign of any structure at the FSN values of $\vec{q}_{CDW}(FSN)$. Figure 2D shows the STM image taken at 22K (below $T_{CDW}$), with the Fourier transform shown in the inset (19). The arrow in the inset marks the position of the periodicity of electronic CDW that is the same as seen in the lattice distortion measured by diffraction (13, 14). Figure 2E is the STM image at 96K (above $T_{CDW}$), whose Fourier transform is missing the spot indicated in the inset of Fig. 3D. The structural transition order parameter is shown in Fig. 2F with the critical temperature at $T_{CDW} = 33.5$ K (17). Fig. 2G displays the measured resistivity as a function of temperatures, clearly showing the superconducting phase transition at $T_S$ but almost no change at the CDW transition temperature $T_{CDW}$ (21). There is no gap in the electronic system, *i.e.*, no insulating state associated with the low temperature CDW state, inconsistent with that predicted by the Peierls model.

Inelastic X-ray scattering reveals what appears to be a temperature dependent Kohn anomaly in the acoustic phonon branch at $\vec{q}_{CDW}$ (20). The data of the renormalized phonon dispersion along the Γ-M direction (Fig. 3A) (20) show the strong phonon softening located around $q_{CDW} \cong 0.695 \text{Å}^{-1}$, indicating the one-to-one correspondence between the CDW and the Kohn-like anomaly. Just below the transition temperature the phonon energy is imaginary, meaning there is a restructuring of the lattice. Above the transition temperature the phonon energy drops near $\vec{q}_{CDW}$ but does not go to zero. Figure 3B displays the phonon width as a function of $q$ in the CDW phase at 33 K, showing an appreciable increase in width in the region of the phonon anomaly (20). This is a signature of the phonon coupling with other excitations in the system (EPC). Comparison of Fig. 3A with Fig. 1B shows that this is not a simple Kohn anomaly, since the phonon is damped over a considerable range of the BZ, in contrast to the sharp dip at $2k_F$ expected from FSN and the Peierls picture (20). The authors show by using *ab initio* calculations that the $\vec{q}$-dependence of this phonon softening is dictated by EPC, not by FSN (20, 24), but they report little about the characteristics of the EPC coupling matrix element $|g(\vec{k},\vec{k}')|^2$, presumably because they did not realize at the time that $|g(\vec{k},\vec{k}')|^2$ could be measured experimentally. But it can be determined experimentally as we will demonstrate, first for NbSe$_2$ and then for optimally doped cuprate Bi$_2$Sr$_2$CaCu$_2$O$_{8+\delta}$ (Bi2212).

The $\vec{q}$-dependent EPC is described by the matrix element $|g(\vec{k},\vec{k}')|^2$ that couples electronic states $\vec{k}$ and $\vec{k}'$ with a phonon of wave vector $\vec{q}$ ($\vec{q} = \vec{k} - \vec{k}'$) and energy $\hbar\omega$ (25). If we make the assumption that $|g(\vec{k},\vec{k}')|^2 = |g(\vec{q})|^2$, following Grimwall (26), we can determine $|g(\vec{q})|^2$ knowing $\text{Im}[\chi(q,\omega)]$ (8),

$$\Gamma_{EPC}(\vec{q}) = -2|g(\vec{q})|^2 \text{Im}[\chi(\omega,\vec{q})], \tag{1}$$

where $\Gamma_{EPC}(\vec{q})$ is the phonon linewidth. Ideally the interacting susceptibility $\text{Im}[\chi(q,\omega)]$, which carries the information from both electron-phonon and electron-electron interactions, should be used for the specific system being studied. However, experimentally it is not feasible to obtain the interacting $\text{Im}[\chi(q,\omega)]$. Thus we have adapted an experimental approach, using the measured band structure which contains both electron-phonon and electron-electron coupling as the input to calculate the Lindhard response function $\text{Im}[\chi_0(q,\omega)]$. In the current case for an acoustic phonon, a good approximation is to use the zero energy value of the Lindhard function $\text{Im}[\chi_0(\omega,q)] = \text{Im}[\chi_0(0,q)] = \text{Im}[\chi_0(q)]$, which is plotted in Fig. 2C. The density-functional-theory calculation in Ref. (6) for the same system, including exchange-correlation interaction between electrons, obtained a very similar response function as shown from our procedure in Fig. 6 of Ref. (6). This indicates that our experimental procedure is a reliable procedure for determining the response function. Figure 3C shows the result of applying Eqn. 1 using the best fit to $\Gamma_{EPC}(q)$ in the Γ-M direction (Fig. 3B, also see supplemental material). It is clear, as anticipated, that the strong Kohn-like anomaly observed in 2H-NbSe$_2$ arises from the $\vec{q}$-dependent EPC matrix element with a peak at $\vec{q}_{CDW}$, not from the conventional FSN in the electronic structure. A consistency check can be performed by using the extracted $|g(\vec{q})|^2$ in the equation derived by Grimvall (25) relating the measured phonon dispersion $\omega(\vec{q})$ to the bare phonon dispersion $\omega_0(\vec{q})$ (without EPC):

$$\omega(\vec{q})^2 = \omega_0(\vec{q})^2 + 2\omega_0(\vec{q})|g(\vec{q})|^2 \text{Re}[\chi(\omega,\vec{q})]. \qquad (2)$$

Replacing $\text{Re}[\chi(\omega,\vec{q})]$ with the zero frequency limit $\text{Re}[\chi_0(\vec{q})]$ plotted in Fig. 2B gives the result displayed in Fig. 3D, where we have approximated $\omega_0(\vec{q})$ by extrapolating the high temperature data shown in Fig. 3A. The predicted acoustic phonon dispersion (solid lines) is in remarkable agreement with the data (20). This example illustrates how important the measurements of phonon dispersion and linewidth are in unraveling the origin of the CDWs in the metallic crystals. However, there are still questions remaining, for example, will a large $|g(\vec{q})|^2$ always lead to the formation of a CDW state with lattice distortion? If not, why?

To answer this question we examine systems with large EPC but very different electronic behavior, such as cuprates. The fact that cuprates belong to the class of unconventional superconductors is not immediately relevant to this study. But in cuprates, different charge ordering (CO) phenomena have been reported, such as a stripe phase in La$_{1.6-x}$Nd$_{0.4}$Sr$_x$CuO$_4$ (26, 27), a checker-board phase (28, 29) and a special kind of short-range CO phase (30, 31) in Bi2212, and long-range charge fluctuations (32) or CDWs (33, 34) in YBa$_2$Cu$_3$O$_{6+x}$ (YBCO). The competition between these CO phases and magnetic ordering and superconductivity (35) continues to attract intensive interest in the community. Comin *et al.* pointed out that the simple FSN near antinodal region is not the driving force of the CO (36). Recent phonon measurements (37, 38) show signatures of strong EPC in the low temperature (below 150 K) charge ordering phase of YBCO. But the role of EPC in YBCO is substantially different than the role in NbSe$_2$ due to two aspects: (1) no phonon softening was observed above 150 K ($T_{CDW}$); (2) the energy of the soft phonon modes never approaches zero. Thus EPC is not strong enough in cuprates to drive the formation of CO or CDW phase. Unlike the case in 2D layered materials like NbSe$_2$, neither FSN nor EPC is directly relevant to the CO or CDW phases in cuprates. Actually a recent STM study (39) has clearly demonstrated that the observed modulations of the electronic structure in cuprates are unconventional density waves, where antiferromagnetic and Coulomb interactions may play a key role. In any case whatever the CO phase in cuprates is, it is not a conventional CDW.

We use optimally doped Bi2212, a typical quasi-2D cuprate with large EPC, as an example, to illustrate the different role of EPC in cuprates. Here a whole set of data exist, specifically for the apical oxygen $A_{1g}$ out of plane optical phonon mode (8). Figure 4A plots our fit to the measured dispersion of the optical phonon branch at ~80 meV (8), showing a rapid downward dispersion for momentum larger than ~ 0.41 Å$^{-1}$, especially in the $(0,0) \to (\pi/a_0, \pi/a_0)$ direction ($a_0 = 3.82$Å, denotes the in-plane lattice constant of Bi2212). Our fit to the linewidth data (9), shown in Fig. 4B, reveals that this downward dispersion is a result of many-body coupling in this system, *i.e.* the linewidth increases dramatically in the regions of large downward dispersion. Figure 4C shows the EPC matrix element $|g(\vec{q})|^2$ of this mode calculated by using published ARPES data that determines $\chi(\vec{q},\omega)$ (40) and the fit to the reported phonon dispersion and linewidth (8). Different from the case of NbSe$_2$, the peak in $|g(\vec{q})|^2$ of Bi2212 is at the zone boundary $(\pi/a_0, \pi/a_0)$, which if large enough in magnitude would have driven the phonon energy imaginary, resulting in a commensurate reconstruction, *i.e.*, a $(\sqrt{2} \times \sqrt{2})R45°$ superstructure which has been reported for Sr$_2$RuO$_4$ when a phonon goes soft at the zone boundary (41). Notice that the maximum in $|g(\vec{q})|^2$ is ~50x10$^3$ meV$^2$Å$^2$ for both NbSe$_2$ (Fig. 3C) and Bi2212 (Fig. 4C).

Knowing $|g(\vec{q})|^2$, can we answer the question why there is a CDW associated with Kohn-like anomaly in NbSe$_2$ but not with the $A_{1g}$ optical phonon mode in Bi2212? When the phonon mode is imaginary at $\vec{q}_{CDW}$ [$\omega^2(\vec{q}_{CDW}) \leq 0$], Eqn. 2 gives

$$|g(\vec{q}_{CDW})|^2 \cong \frac{\omega_0(\vec{q}_{CDW})}{2\left|\text{Re}\left[\chi(\vec{q}_{CDW},\omega)\right]\right|}. \tag{3}$$

For NbSe$_2$ the critical value of $|g(\vec{q})|^2$ is approximately given by $|g_{crit}(\vec{q}_{CDW})|^2 \approx \frac{15\text{meV}}{2(0.16\text{eV}^{-1} \cdot \text{Å}^{-2})} = 48 \times 10^3 \text{meV}^2 \cdot \text{Å}^2$, but for Bi2212 it is $|g_{crit}(\vec{q}_{CDW})|^2 \approx \frac{80\text{meV}}{2(0.25\text{eV}^{-1} \cdot \text{Å}^{-2})} = 160 \times 10^3 \text{meV}^2 \cdot \text{Å}^2$. Obviously the oxygen $A_{1g}$ phonon mode in Bi2212 does not go to zero because the energy is too large. If there is an incipient CDW instability in a cuprate that is associated with EPC then there must be a low energy phonon whose energy becomes imaginary (32).

Why is the shape of EPC matrix element in Bi2212 so different from that in NbSe$_2$? A full understanding will only come when we understand the microscopic picture of the coupling between specific phonons and electronic states of the system, *i.e.* which states contribute to $|g(\vec{q})|^2$ at any $\vec{q}$? To evaluate the contributions from the electronic bands to $|g(\vec{q})|^2$ in Bi2212, we plot in Fig. 4D both $\text{Re}[\chi(\vec{q},0)]$ and $\text{Re}[\chi(\vec{q},80\text{meV})]$ in the $(0,0) \rightarrow (\pi/a_0, \pi/a_0)$ direction, illustrating their dramatic difference with energy. One might have argued from $\text{Re}[\chi(\vec{q},0)]$ that this system would support a CDW at $q\sim0.41$ Å$^{-1}$ because of the peak at this momentum, but there is no peak in $\text{Re}[\chi(\vec{q},80meV)]$ at this momentum. The inset in Fig. 4D shows $\text{Im}[\chi(\vec{q},80meV)]$ with a peak at $q\sim0.18$ Å$^{-1}$, which is also present in $\text{Re}[\chi(\vec{q},80\text{meV})]$, indicating that there could be a Kohn-like anomaly in the A$_{1g}$ phonon branch at this $q$. But in fact nothing is seen in the phonon dispersion or linewidth at this small $q$ (see Fig. 4 A&B). Here again it is illustrated that although the shape of the susceptibility function is important for determining the characteristics of $|g(\vec{k},\vec{k}')|^2$, it appears to have little predictive power when it comes to predicting the existence of or the characteristic of a CDW (4).

To gain insight into the microscopic picture of EPC for this specific phonon, we must learn which states contribute to $|g(\vec{q})|^2$ at a given $\vec{q}$. In Fig. 4E we show the contour plot of the electronic band within the phonon energy range near Fermi energy (0±80 meV) (40). Note how slowly the band disperses in the $(0,0) \rightarrow (\pi/a_0, 0)$ direction near the BZ boundary, in contrast to that in the $(0,0) \rightarrow (\pi/a_0, \pi/a_0)$ direction. The initial electron states contributing to the susceptibility functions will be mainly located in this region of k-space. This is why the susceptibility functions shown in Fig. 4D is so dependent upon ω, and obviously influences the shape of $|g(\vec{q})|^2$. To illustrate this, we select the momentum $\vec{q}=(\pi/a_0, \pi/a_0)=(0.822\text{Å}^{-1},0.822\text{Å}^{-1})$ where $|g(\vec{q})|^2$ has its maximum value (Fig. 4C), and plot the relative density of initial states contributing to $|g(\vec{q})|^2$ (see supplemental material) in Fig. 4F. Allowed initial states along the vector $\vec{k}=(k_x,0)$ are only allowed within a very narrow range of momentum and energy (conservation of energy and momentum), from $k_x=0.632\text{Å}^{-1}$ to $0.638\text{Å}^{-1}$, and 74.4 meV to 79.7 meV, respectively. Because of the flat band characteristic in the $(0,0) \rightarrow (\pi/a_0, 0)$ direction (Fig. 4E) these allowed initial states are indeed far from the Fermi Contour. In the supplemental material we show the plot for $\vec{q}=(\pi/a_0,0)=(0.822\text{Å}^{-1},0)$, which is dramatically different from that in Fig. 4F.

In summary, we have illustrated in this paper how to experimentally determine the details contained in the EPC matrix element by combining measurements of the electron band dispersion with the phonon dispersion and linewidth. The contribution of specific electronic and vibrational states can be identified. For those materials where EPC dictates the CDW wave vector, this procedure should lead to a new understanding of the origin and the effect of CDWs in solids. There are two important additional questions yet to be answered (especially from the aspect of theoretical modeling) – what determines the characteristics of EPC matrix element, and why? It is worth pointing out that the proposed procedure will not work for all systems, because of the basic physics associated with different classes of CDW. The concept of a CDW has been applied to many materials without a clear definition of the essence of CDW. We suggest that there are at least three types of CDWs. Type I CDWs are quasi-1D systems with their origin in the Peierls instability (FSN). Here the lattice distortion is a secondary effect (4) and results from the electronic disturbance. The best example may be linear chain compounds (1). Type II CDWs are driven by EPC but not by FSN. Here the electronic and lattice instabilities are intimately tied to each other, and there is a phonon mode at $\vec{q}_{CDW}$ going to zero at the transition temperature $T_{CDW}$. There is no reason to have a metal-insulator transition associated with the transition. Type II CDWs have been described in this paper with NbSe$_2$ as the example. Type III CDWs are systems where there is a charge modulation (or CO) with no indication of FSN or EPC as the driving force. The best example of Type III CDWs may be the cuprates that exhibit CO phenomena (26-39). Strong EPC or FSN may exist in those systems, but there is no clear signature that EPC or FSN are essential to the formation of the CDWs. We emphasize that measurement of phonon dispersion and linewidth coupled with electron spectroscopy is the way to classify the types of CDWs and to understand their origin. This is especially true for Type III systems where the origin of or even the existence of CDWs is still under debate (42).

**Acknowledgements:** We would like to thank Drs. John Tranquada and Igor Mazin for their comments on this manuscript. We would also like to thank Drs. C. J. Arguello, S. P. Chockalingam and A. N. Pasupathyfor sharing their STM images of NbSe$_2$. The research was funded by 973 Project (2012CB921700), NSF (11225422 & 11304367) of China, and the External Cooperation Program of BIC, CAS (112111KYSB20130007).

**Author contributions:** XZ and YC contributed equally to the data analysis. All of the authors contributed equally to the manuscript preparation.

**Figure Captions:**

**Figure 1:** Schematic representations of the physical phenomena associated with Fermi Surface nesting. **A**: The plot of the one-dimensional free electron band structure for a chain of atoms with separation $a$ and one electron per atom site. The location of the Fermi Energy is indicated with the nesting vector $q$. **B**: Kohn anomaly in the acoustic phonon branch as a function of temperature, which results from Fermi Surface nesting. **C**: The Fermi contour of a two-dimensional free electron system. **D**: Lindhard Response Function $\chi_0(\vec{q})$ as a function of dimensionality (1). The dotted line shows the effect of 2% deviation from perfect FSN and the inset is for a small "Drude Relaxation" (4).

**Figure 2:** Properties of the CDW in NbSe$_2$. **A**: Fermi Surface Contours (10) with the experimental $\vec{q}_{CDW}$ (16, 17) shown. **B** and **C**: Re $\chi_0(q)$ and Im $\chi_0(q)$ calculated from experimental data (10) along the ΓM direction. **D**: STM image of the CDW phase at $T=22$ K. The inset shows the Fast Fourier Transform (19). **E**: STM image at a temperatures above $T_{CDW}$ (19). **F**: Order parameter of the lattice transition associated with the CDW (17). **G**: Resistivity as a function of temperature, with the superconducting and CDW transition temperatures indicated (21). **D** and **E** are adapted with permission from Ref. (19), copyrighted by the American Physical Society.

**Figure 3:** Acoustic phonon softening in NbSe$_2$ along the ΓM direction. **A**: Measured phonon dispersion as a function of $T$ (20). **B**: Measured linewidth of the phonon at $T=33K$ (20). **C**: Extracted EPC matrix element $|g(\vec{q})|^2$ (see text and Eqn. 1). **D**: Calculated soft phonon behavior (solid line) using $|g(\vec{q})|^2$ and an approximation for the bare phonon (dashed line) dispersion.

**Figure 4:** EPC associated with apical oxygen A$_{1g}$ phonon mode in optimally doped Bi$_2$Sr$_2$CaCu$_2$O$_{8+\delta}$. **A**: Fit to the measured phonon dispersion (8). **B**: Fit to the measured linewidth as a function of momentum (8). **C**: $|g(\vec{q})|^2$ extracted from the phonon and ARPES data (8, 40). **D**: Calculated Re $\chi(\vec{q},\omega)$ for ω=0 & 80 meV from the ARPES data along $(0,0) \rightarrow (\pi/a_0, \pi/a_0)$ direction. The inset is the calculated Im $\chi(\vec{q},\omega)$ for 80 meV. **E**: Constant energy contours within 80 meV of the Fermi surface (40). **F**: Density of possible initial states $\vec{k}$ calculated with $|g(\vec{q})|^2$ in **C** where the final states $\vec{k}'$ are above the Fermi energy and the initial and final electronic states are separated by $\vec{q} = (\pi/a_0, \pi/a_0) = (0.822\text{Å}^{-1}, 0.822\text{Å}^{-1})$ (white arrow) with the phonon energy $\hbar\omega$.

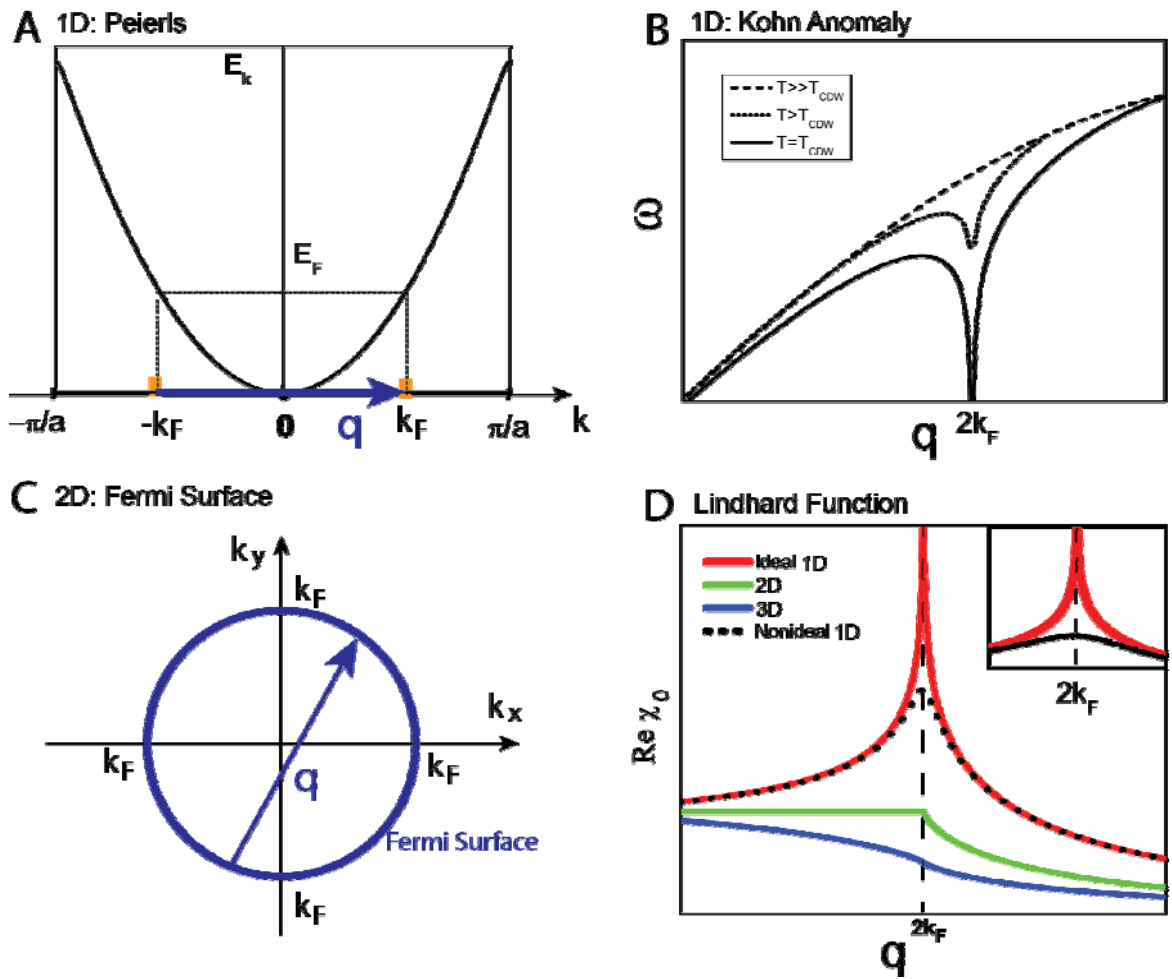

**Zhu** *et al.* **Fig. 1**

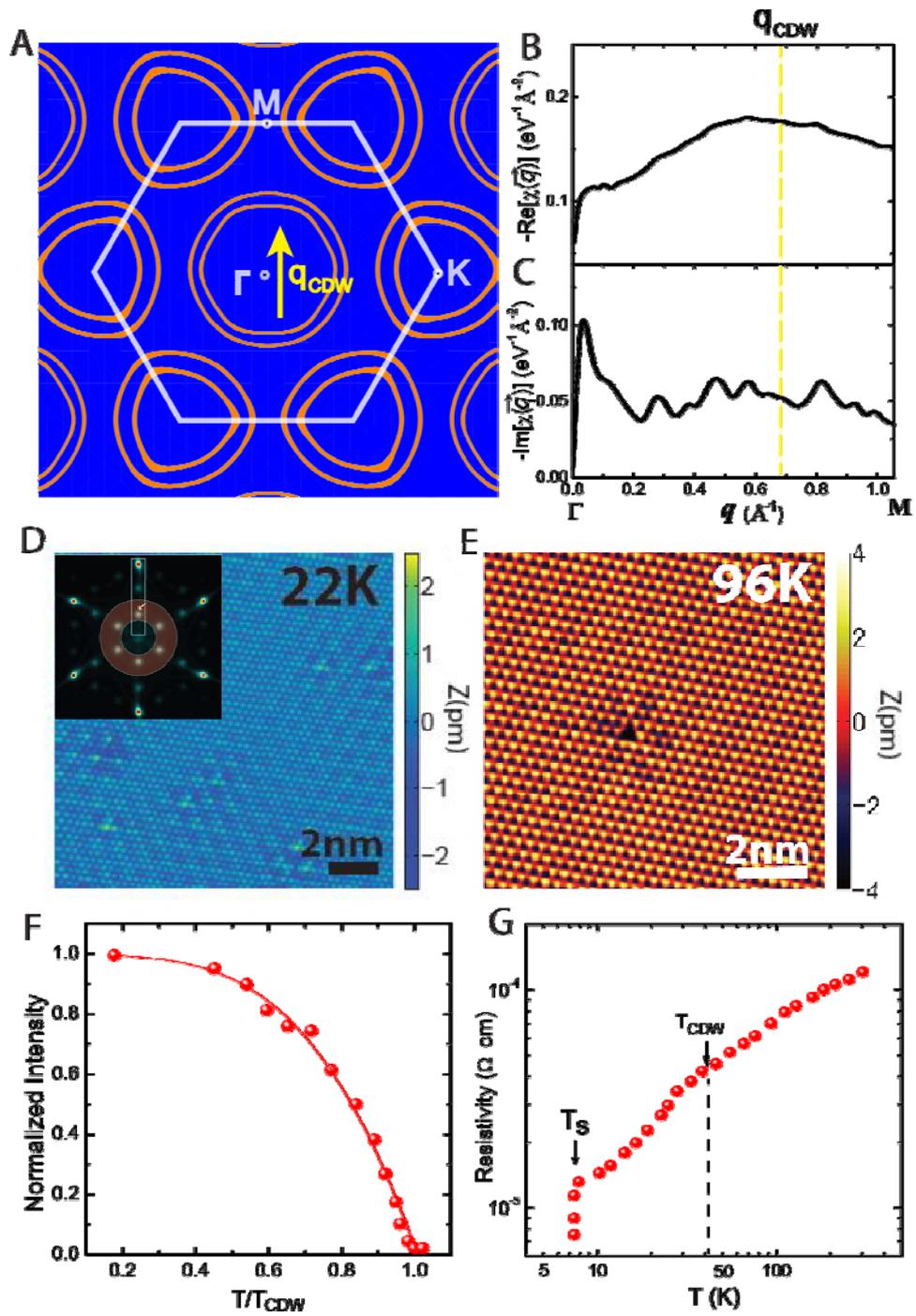

**Zhu** *et al.* **Fig. 2**

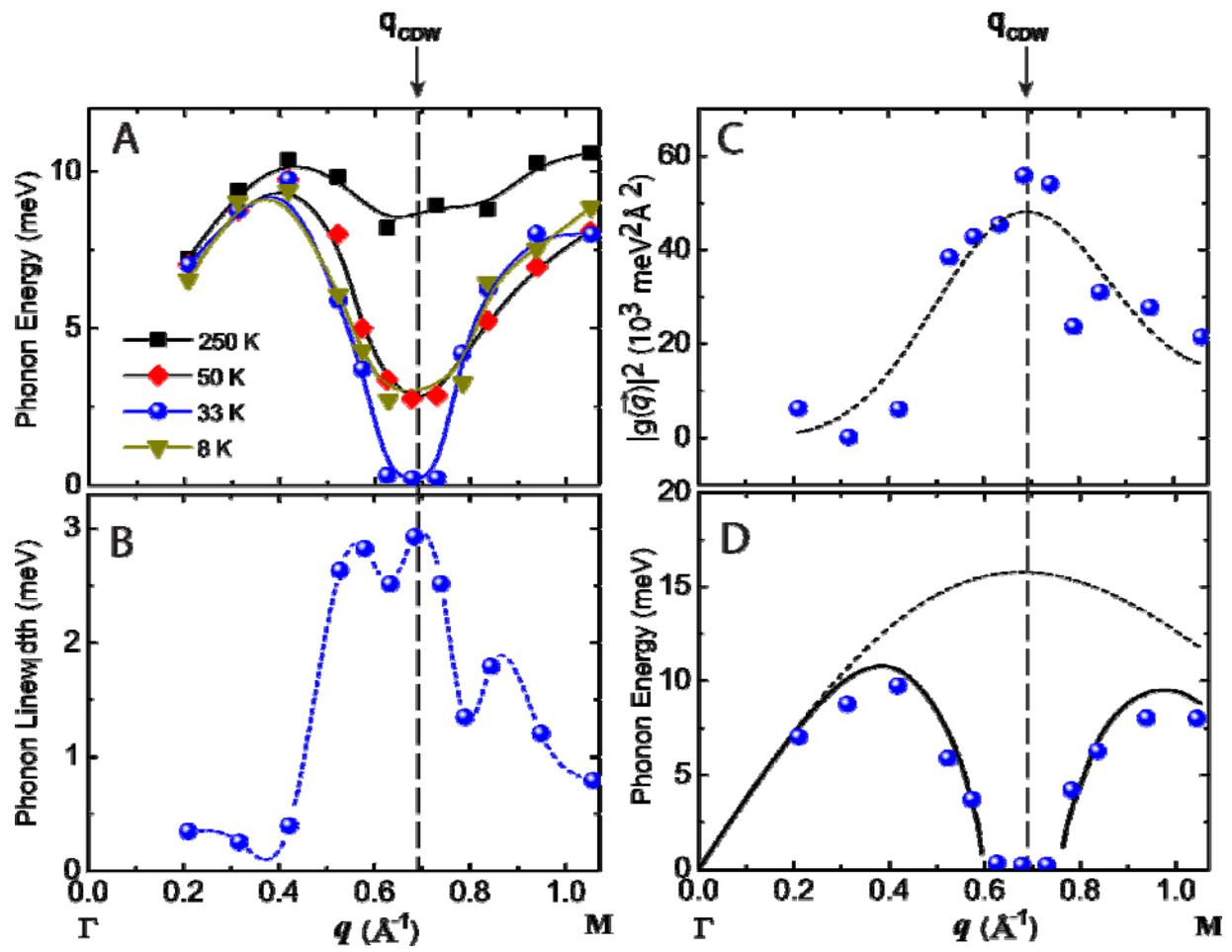

**Zhu** *et al.* **Fig. 3**

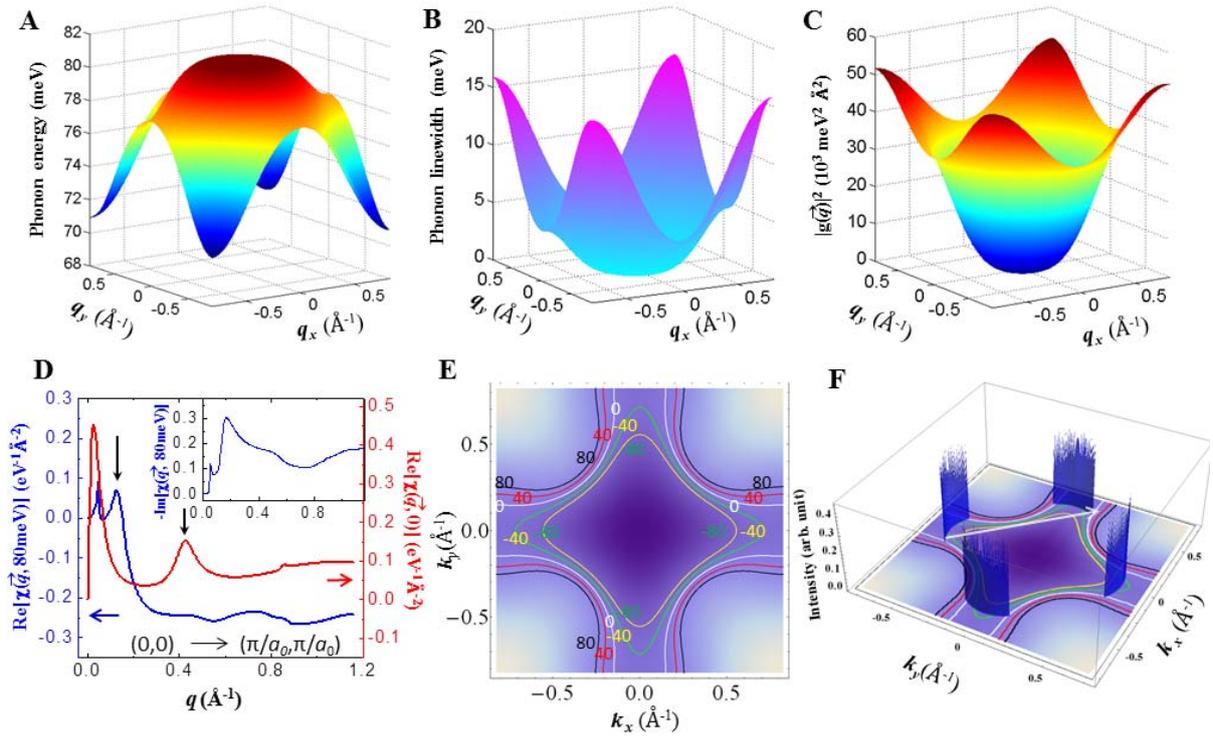

**Zhu *et al.* Fig. 4**

11-03-2014

# Supporting Information
## Classification of Charge Density Waves Based on Their Nature

### 1. The susceptibility function

In the random phase approximation (RPA), the susceptibility function is expressed as

$$\chi(\vec{q},\omega) = -\frac{2iV}{(2\pi)^3}\int_{BZ} d\vec{k}\, G^0(\varepsilon_0,\vec{k}) G^0(\varepsilon_0+\omega,\vec{k}+\vec{q}),  \tag{S1}$$

where $G^0$ is the electron Green's function, $\varepsilon_0$ is the electron's energy and $\omega$ is the phonon's energy, respectively. In the noninteracting Fermi Gas model, electron-electron correlations are ignored and the susceptibility function can be calculated via Lindhard approximation. The transition momentum is conserved with $\vec{k}+\vec{q}=\vec{k}'$, assuming crystal volume $V\equiv 1$, the imaginary and real parts of the Lindhard susceptibility function $\chi_0(\vec{q},\omega)$ are:

$$\mathrm{Im}[\chi_0(\vec{q},\omega)] = \frac{1}{2\pi}\int_{BZ} d\vec{k}(f_{\vec{k}'}-f_{\vec{k}})\delta(\omega+\varepsilon_{\vec{k}}-\varepsilon_{\vec{k}'}), \tag{S2}$$

and

$$\mathrm{Re}[\chi_0(\vec{q},\omega)] = \frac{2}{(2\pi)^2}\int_{BZ} d\vec{k}\,\frac{f_{\vec{k}'}-f_{\vec{k}}}{\omega+\varepsilon_{\vec{k}}-\varepsilon_{\vec{k}'}}, \tag{S3}$$

where $f$ denotes the electron's Fermi distribution function.

For the special case of zero-energy transfer limit, $\omega=0$, and $\mathrm{Im}[\chi_0(\vec{q},0)]=0$. The Lindhard functions as a function of dimensionality are:

$$\textbf{3D:}\quad \mathrm{Re}\,\chi_0 \sim [1+\frac{1-(q/2)^2}{q}\ln\left|\frac{1+q/2}{1-q/2}\right|] \tag{S4}$$

$$\textbf{2D:}\quad \mathrm{Re}\,\chi_0 \sim \begin{cases}(1-\sqrt{1-(2/q)^2}), q\geq 2k_F \\ 1/E_F, q<2k_F\end{cases} \tag{S5}$$

$$\textbf{1D:}\quad \mathrm{Re}\,\chi_0 \sim \frac{1}{2q}\ln\left|\frac{1+q/2}{1-q/2}\right| \tag{S6}$$

These results are plotted in Figure 1D in the paper.

## 2. Fitting of the EPC matrix element for NbSe2

The EPC matrix element for NbSe$_2$ determined from the phonon linewidth data (1) is fitted, using the least-squares method, by the following expression:

$$|g(\vec{q})|^2 = \{3(a+b+c+d+...) - a[\cos(q_x a_0) + 2\cos(\frac{1}{2}q_x a_0)\cos(\frac{\sqrt{3}}{2}q_y a_0)]$$
$$-b[\cos(\sqrt{3}q_y a_0) + 2\cos(\frac{3}{2}q_x a_0)\cos(\frac{\sqrt{3}}{2}q_y a_0)] \quad \text{(S7)}$$
$$-c[\cos(2q_x a_0) + 2\cos(q_x a_0)\cos(\sqrt{3}q_y a_0)]$$
$$-d[\cos(2\sqrt{3}q_y a_0) + 2\cos(3q_x a_0)\cos(\sqrt{3}q_y a_0)] - ...\}^2$$

which has a form of a Fourier expansion for a lattice with hexagonal symmetry. Since the phonon data are only available along the Γ-M direction, the above two-dimensional $|g(\vec{q})|^2$ was reduced to one-dimension by setting $q_x = 0$, giving

$$F1 = [A + B + C - A\cos(\frac{\sqrt{3}}{2}q_y a_0) - B\cos(\sqrt{3}q_y a_0) - C\cos(\frac{3\sqrt{3}}{2}q_y a_0)] \quad \text{(S8)}$$

an approximation to the third order, and

$$F2 = [A + B + C + D - A\cos(\frac{\sqrt{3}}{2}q_y a_0) - B\cos(\sqrt{3}q_y a_0)$$
$$-C\cos(\frac{3\sqrt{3}}{2}q_y a_0) - D\cos(2\sqrt{3}q_y a_0)] \quad \text{(S9)}$$

an approximation to the fourth order. The fitting of $|g(\vec{q})|^2$ using these two functions are shown in Fig. S1.

A consistency check of the phonon dispersion can be performed (via Eqn. (2) in the paper) by using the fitted $|g(\vec{q})|^2$ from the above two functions. In this calculation, $\omega_0(\vec{q})$ was approximated by extrapolating the high temperature data as a sinusoidal function, which is a reasonable approximation for the acoustic phonon mode. The results are plotted in Fig. S2.

In the paper, only the results fitted by function F1 are displayed (Fig. 3 C&D). Actually both of the two fittings show fairly good agreement with the experimental data, indicating that the data quality is not good enough. This is a sign that better and denser phonon linewidth data are needed. And it also implies the importance of the phonon linewidth measurements to determine the EPC matrix element.

## 3. Fitting of the phonon measurements for Bi2212

The phonon data (both energy and linewidth) for the apical oxygen $A_{1g}$ mode of Bi2212 obtained by angle-resolved high-resolution electron energy loss spectroscopy (2) is shown in Fig. S3. There is no discernable difference between the data obtained at 60 K and those obtained at room temperature (RT). This demonstrates that contributions to the linewidth from phonon-phonon and phonon-defects interactions can be ignored, since those interactions are strongly temperature-dependent, justifying considering only EPC in our calculation. The measured dispersion and linewidth are fitting by the following functions:

$$\omega(\vec{q}) = 80.7 - \{1.5636 - 0.8362[\cos(q_x a_0) + \cos(q_y a_0)] + 0.1088\cos(q_x a_0)\cos(q_y a_0)\}^2, \text{ (S10)}$$

and

$$\gamma(\vec{q}) = \{1.9888 - 1.0433[\cos(q_x a_0) + \cos(q_y a_0)] - 0.0978\cos(q_x a_0)\cos(q_y a_0)\}^2 \text{ (S11)}$$

These results are plotted in Fig. 4 A&B, respectively, in the paper.

## 4. Details of the states contributing to EPC matrix element in Bi2212

The relative density of initial states contributing to $|g(\vec{q})|^2$ in Bi2212 is calculated by considering the conservation of energy and momentum of the electron transition from $\vec{k}$ to $\vec{k}'$ via a phonon with energy $\hbar\omega$. In this calculation the initial states were confined to be within 80 meV below the Fermi level, and then the final states would also be within 80 meV above the Fermi level. We select the momentum $\vec{q} = (\pi/a_0, \pi/a_0) = (0.822\text{Å}^{-1}, 0.822\text{Å}^{-1})$, where $|g(\vec{q})|^2$ has its the maximum value (Fig. 4C in the paper), and plot the relative density of initial states contributing to $|g(\vec{q})|^2$ in Fig. S4A. The allowed initial states are confined in four thin "arcs" in the $k_x$-$k_y$ plane. In order to visualize the details of the initial state distribution, half of the arc is projected in the $k_x$-$k_y$ plane and shown in Fig. S4B. It's easy to see that the initial states along the vector $\vec{k} = (k_x, 0)$ are only allowed within a very narrow range of momentum, from $k_x = 0.632\text{Å}^{-1}$ to $0.638\text{Å}^{-1}$. The corresponding transfer phonon energy are from 74.4 meV to 79.7 meV. The relative intensity along this line ( $k_x = 0.632 \text{Å}^{-1}$ to $0.638 \text{Å}^{-1}$ ) is plotted in Fig. S4B.

We also plot the relative intensity of the initial states for $\vec{q} = (\pi/a_0, 0) = (0.822 \text{Å}^{-1}, 0)$ in Fig. S5, which is dramatically different than that in Fig. S4. The allowed states are restricted to a much smaller regions than those for $\vec{q} = (\pi/a_0, \pi/a_0) = (0.822 \text{Å}^{-1}, 0.822 \text{Å}^{-1})$, because of the shapes of the phonon and electronic dispersion. This result explains why the value of $|g(\vec{q})|^2$ at $(\pi/a_0, \pi/a_0)$ is lower than that at $(\pi/a_0, 0)$.

**Supporting Reference**

1. Weber F, *et al.* (2011) Extended phonon collapse and the origin of the charge-density wave in 2H-NbSe$_2$. *Phys Rev Lett* 107(10):107403.
2. Qin HJ, *et al.* (2010) Direct determination of the electron-phonon coupling matrix element in a correlated system. *Phys Rev Lett* 105(25):256402.

**Supporting Figure Captions:**

**Figure S1** Fitting of $|g(\vec{q})|^2$ for NbSe$_2$ using two different functions (details in the text).

**Figure S2** Consistency check of the phonon dispersion with the fitted $|g(\vec{q})|^2$ from two different functions for NbSe$_2$ (details in the text).

**Figure S3** The phonon data (**A** energy; **B** linewidth) for the apical oxygen A$_{1g}$ mode of Bi2212, along the $(0,0) \to (\pi/a_0, \pi/a_0)$ direction, obtained by angle-resolved high-resolution electron energy loss spectroscopy. Solid circles for 60 K and open circles for room temperature, respectively.

**Figure S4** Details about the density of possible initial states $\vec{k}$ of Bi2212 with $\vec{q} = (\pi/a_0, \pi/a_0) = (0.822 \text{Å}^{-1}, 0.822 \text{Å}^{-1})$. **A** is the same plot as Fig. 4F in the paper. The zoom in of the black rectangular region is projected to the $k_x$-$k_y$ plane as shown in **B**. The intensity along the vector $\vec{k} = (k_x, 0)$ from $k_x = 0.632 \text{Å}^{-1}$ to $0.638 \text{Å}^{-1}$ is plotted in **C**.

**Figure S5** Density of possible initial states $\vec{k}$ of Bi2212 calculated with $|g(\vec{q})|^2$ where the final states $\vec{k}'$ are above the Fermi energy and the initial and final electronic states are separated by $\vec{q} = (\pi/a_0, 0) = (0.822 \text{Å}^{-1}, 0)$ (white arrow) with the phonon energy $\hbar\omega$.

11-03-2014

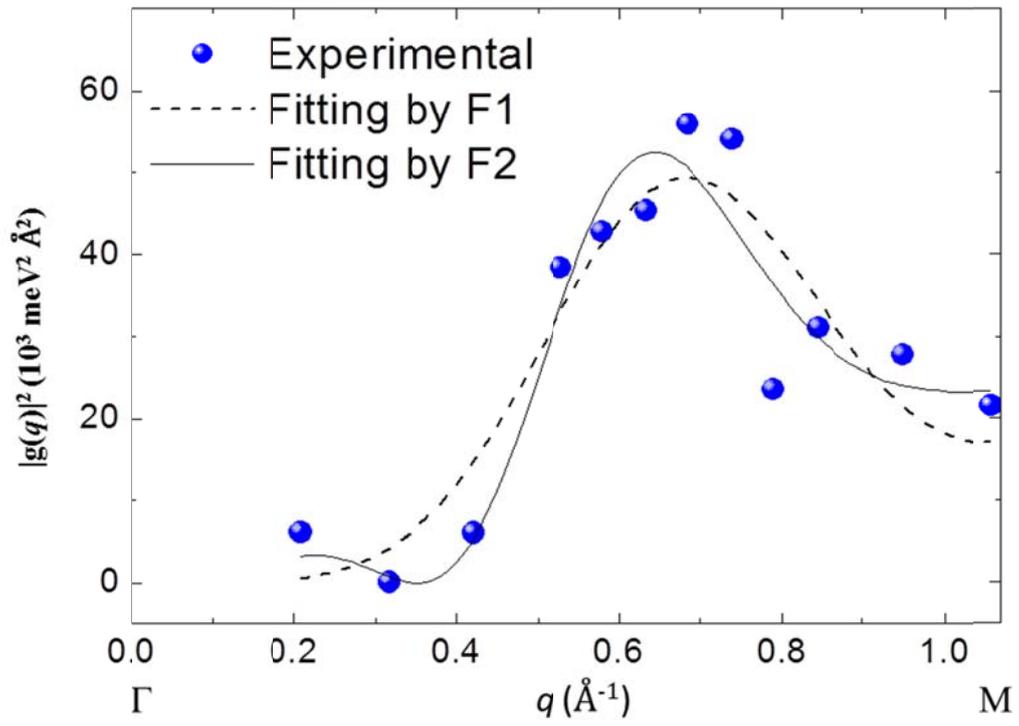

**Zhu *et al.* Fig. S1**

11-03-2014

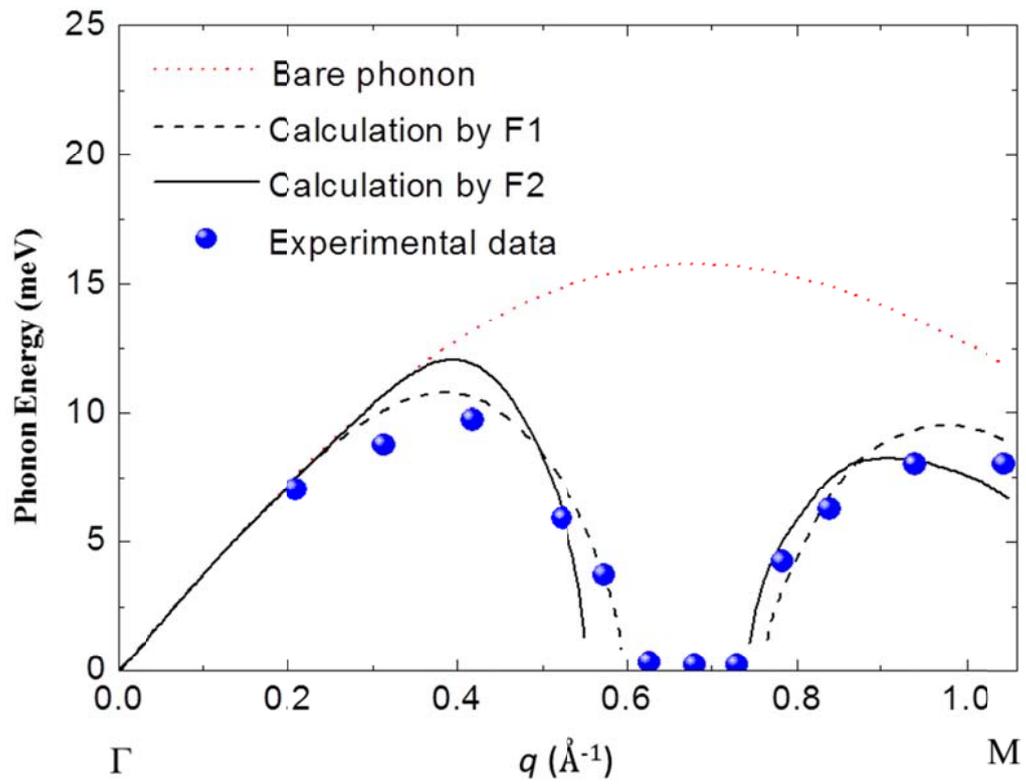

**Zhu** *et al.* **Fig. S2**

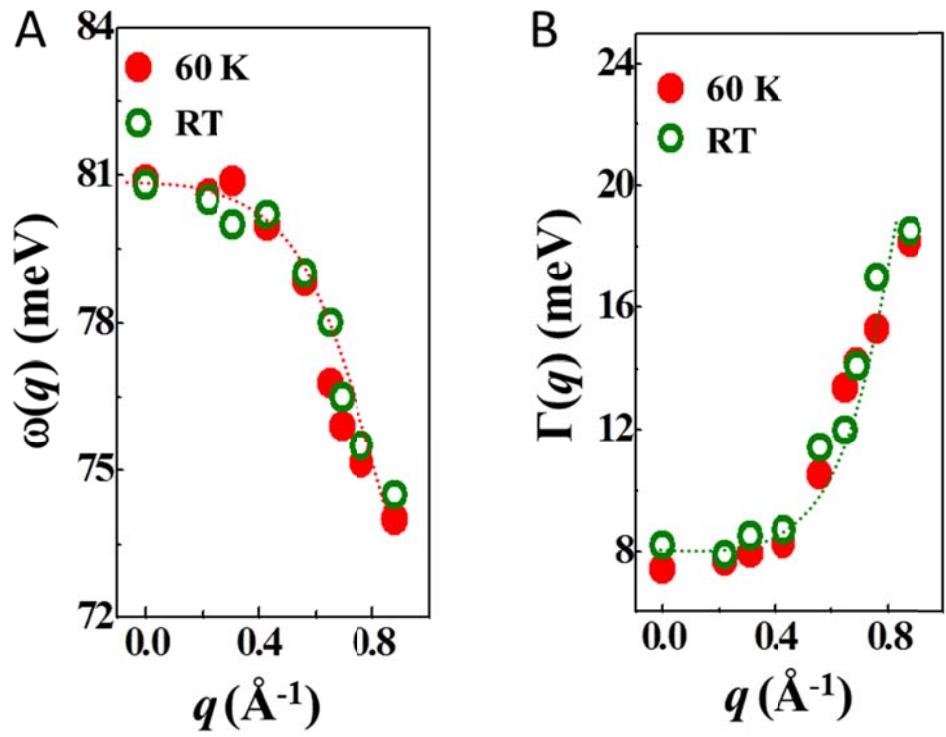

**Zhu *et al.* Fig. S3**

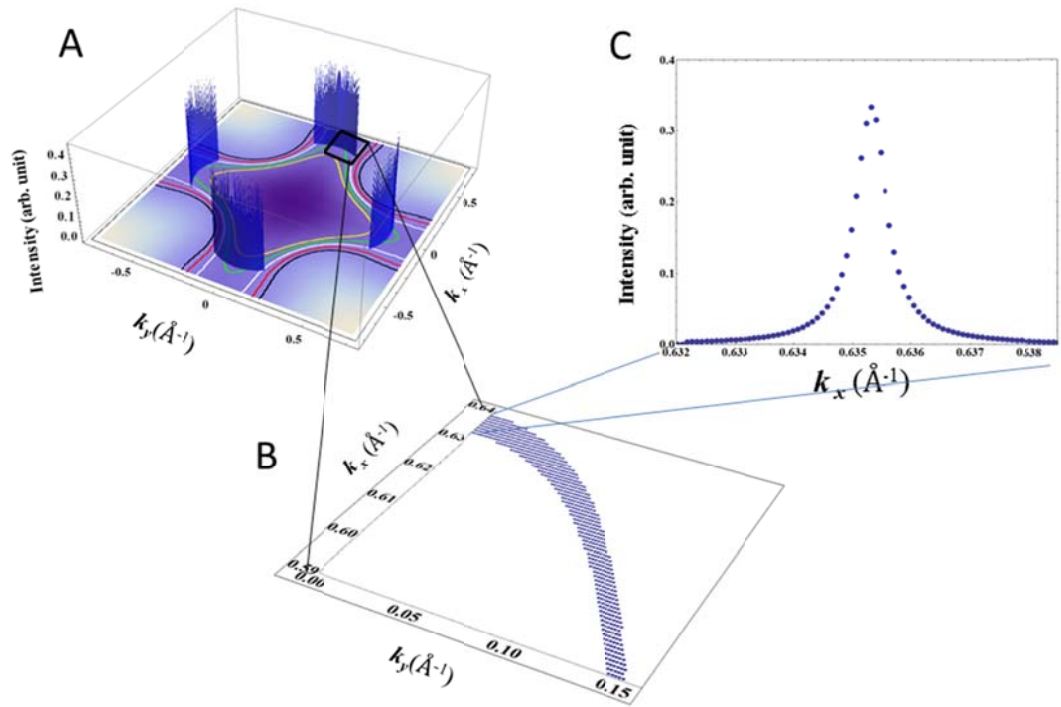

**Zhu** *et al.* **Fig. S4**

11-03-2014

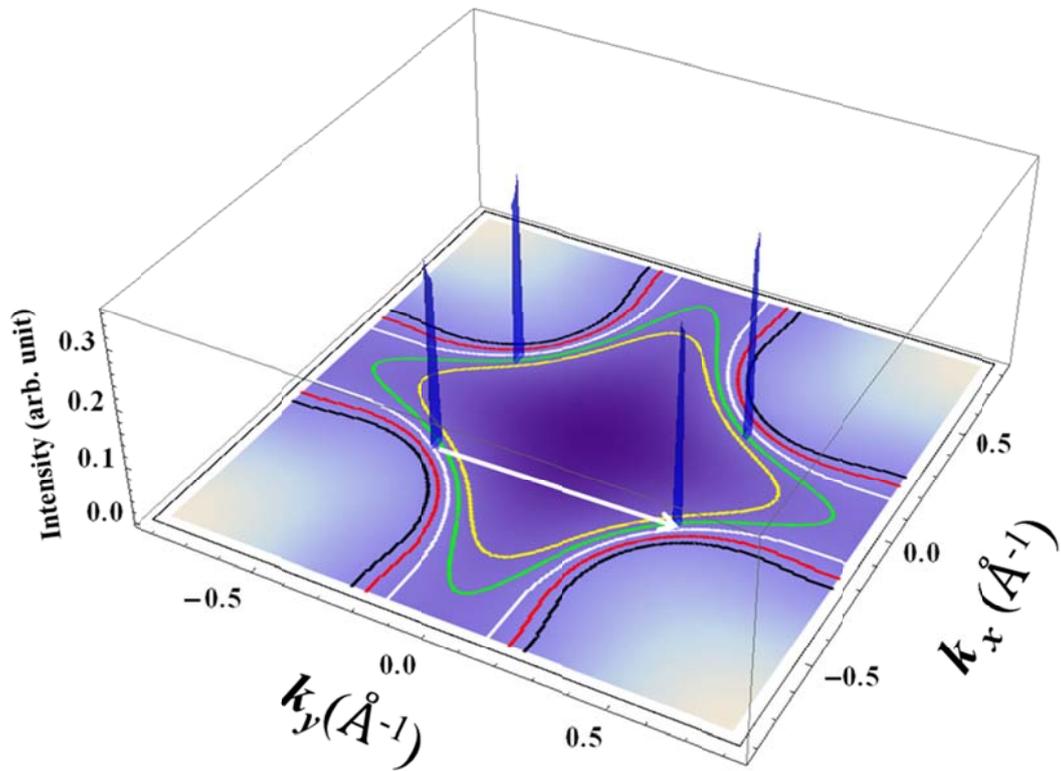

**Zhu *et al.* Fig. S5**